\documentclass[final,3p,times,twocolumn]{elsarticle}

\usepackage{url}                                        
\usepackage{xcolor}                                     
\usepackage{amssymb}                                    
\usepackage{amsmath,amsthm}
\usepackage{amsthm}                                     
\usepackage{epstopdf}                                   
\usepackage[caption=false,font=footnotesize]{subfig}    
\usepackage{float}
\usepackage{placeins}
\usepackage{multirow}                                   
\usepackage{etoolbox}                                   
\usepackage{verbatim}                                   
\usepackage[makeroom]{cancel}                           
\usepackage[utf8]{inputenc}                             
\usepackage{listings}                                   
\usepackage{enumitem}									
\usepackage{xspace}

\newcommand{\fig}[1]     {Fig.~\ref{#1}}
\newcommand{\tab}[1]     {Table~\ref{#1}}

\DeclareMathOperator\erf{erf}

\newcommand{\sobs}       {\ensuremath{\sigma_{\mathrm{obs}}}\xspace}
\newcommand{\sstat}      {\ensuremath{\sigma_{\mathrm{stat}}}\xspace}
\newcommand{\sanis}      {\ensuremath{\sigma_{\mathrm{anis}}}\xspace}

\newcommand{\ALO}        {\ensuremath{A_{\LO}}\xspace}
\newcommand{\ATTT}       {\ensuremath{A_{\TTT}}\xspace}

\newcommand{\LO}         {\ensuremath{L}\xspace}

\newcommand{\TTT}        {\ensuremath{S}\xspace}
\newcommand{\LOave}      {\ensuremath{\hat{\LO}}\xspace}
\newcommand{\TTTave}     {\ensuremath{\hat{\TTT}}\xspace}
\newcommand{\LOaverange} {\ensuremath{\hat{\LO}\pm\sigma}\xspace}

\newcommand{\Erecoil}    {\ensuremath{E_{\textrm{recoil}}}\xspace}

\newcommand{\genunit}[2]{\ensuremath{#1~\text{#2}}\xspace}

\newcommand{\cm}[1]     {\genunit{#1}{cm}}

\newcommand{\MeV}[1]    {\genunit{#1}{MeV}}

\newtoggle{color}
\toggletrue{color} 

\journal{Nuclear Instruments and Methods in Physics A}

\begin{document}

\begin{frontmatter}

\title{Characterization of the scintillation anisotropy in crystalline stilbene scintillator detectors}

\author[ps]{P.~Schuster\corref{cor1}}
\ead{pfschus@berkeley.edu}

\author[eb]{E.~Brubaker}
\ead{ebrubak@sandia.gov}

\cortext[cor1]{Corresponding author}

\address[ps]{Department of Nuclear Engineering, University of California, Berkeley, CA, USA}
\address[eb]{Sandia National Laboratories, Livermore, CA}

\begin{abstract}
This paper reports a series of measurements that characterize the directional dependence of the scintillation response of crystalline melt-grown and solution-grown stilbene to incident DT and DD neutrons. These measurements give the amplitude and pulse shape dependence on the proton recoil direction over one hemisphere of the crystal, confirming and extending previous results in the literature for melt-grown stilbene and providing the first measurements for solution-grown stilbene. In similar measurements of liquid and plastic detectors, no directional dependence was observed, confirming the hypothesis that the anisotropy in stilbene and other organic crystal scintillators is a result of internal effects due to the molecular or crystal structure and not an external effect on the measurement system. 
\end{abstract}


\end{frontmatter}


\section{Introduction}\label{sec:intro}
Stilbene is an organic crystal scintillator that has been used for many decades for radiation detection. Recently, a new solution-based growth method has been developed that produces large crystals with high light output and excellent neutron-gamma pulse shape discrimination (PSD)~\cite{Carman2013,Zaitseva2015,Bourne2016}. This solution-grown stilbene is receiving substantial interest over liquid and plastic alternatives as its performance is better than liquid and PSD-capable plastic scintillators~\cite{Zaitseva2015,Pozzi2013,Cester2014}, and its solid form is often easier to work with than liquid scintillators that are subject to thermal expansion and risk of leaks.

Crystal organic scintillators are known to have a directionally dependent scintillation response for heavy charged particle interactions. The directional dependence of anthracene has been widely characterized by previous authors~\cite{Schuster2016,Brubaker2010,Brooks1974,Tsukada1965}, but only limited measurements have been reported for the directional dependence of stilbene. Previous measurements of proton recoil events in stilbene include the magnitude of change in light output  at 3.7, 8, and \MeV{22}~\cite{Tsukada1962,Brooks1974}, and the light output vs. angle in two planes at \MeV{3.7}~\cite{Tsukada1962} and in one plane at \MeV{14}~\cite{Brubaker2010}. Thus far, no measurements of the effect across a full hemisphere of crystal axes have been provided. Additionally, all previous measurements were of melt-grown stilbene crystals. Given the rising popularity of solution-grown stilbene as a detection material, it is important to thoroughly characterize the directional dependence and understand if it depends on crystal size or growth method.

This paper characterizes the scintillation anisotropy in melt-grown and solution-grown stilbene for \MeV{14.1} and \MeV{2.5} proton recoil events. The magnitude of change in the light output and pulse shape has been quantified, and the light output and pulse shapes have been characterized as a function of direction for a full hemisphere worth of proton recoil directions. In order to investigate how consistent the anisotropy effect is across stilbene detectors, four stilbene samples have been characterized. These include two solution-grown samples with the same dimensions, a solution-grown sample of different dimensions, and a melt-grown sample of different dimensions. 

Characterizing the effect across these four detectors will provide information on whether the effect depends on crystal geometry or growth method, and will evaluate whether the effect varies significantly between two seemingly identical stilbene samples. The geometry of a crystal has been hypothesized to affect the anisotropy as it impacts the light collection efficiency and PSD performance~\cite{Bourne2015}. The growth method and quality of the crystal has also been hypothesized to impact the anisotropy since the growth method does impact the light output and PSD performance~\cite{Carman2013,Zaitseva2015}. 

Previous authors stated that no anisotropy effect was observed in amorphous plastic and liquid materials~\cite[p.~261]{Birks1964}, but no measurements have been published to support that statement. This paper reports measurements of plastic and liquid scintillators that verify no anisotropy is observed, supporting the leading hypothesis that the physical mechanism that produces the anisotropy comes from the molecular or crystal structure within the material and is not caused by an external effect on the measurement system~\cite{Schuster2016}.

\section{Measurements}
In order to characterize the directional dependence in these materials, the detector response to proton recoil events at different directions in the detector was measured. The method for measuring proton recoil events at a fixed energy as a function of recoil direction was the same as in a previous paper which contains a more detailed description than will be presented here~\cite{Schuster2016}. To summarize, monoenergetic neutrons were produced using a Thermo Electric MP 320 DD (D+D$\rightarrow^3$He+$n$; $E_n$=2.5 MeV) or DT (D+T$\rightarrow^4$He+$n$; E$_n$=14.1 MeV) neutron generator. Events in which the neutron deposited approximately its full energy to the recoil proton were selected in order to choose proton recoils with energy $\Erecoil\approx E_n$ traveling in the same direction as the incident neutron. A motor-driven rotational stage was used to position the detector at any angle in $4\pi$ with respect to the incident neutron direction. The detectors were placed approximately \cm{152} from the neutron generator, controlling the incident angle of the neutrons on the detector within 2$^\circ$. 

Events were recorded using a Struck SIS3350 500~MHz 12-bit digitizer. The digitial signal processing technique was the same as described in~\cite{Schuster2016}. For each event the light output \LO and pulse shape parameter \TTT were calculated. \LO was calculated as the sum of digitized samples in each pulse. \TTT was calculated as the fraction of the light in a defined delayed region of the pulse. The boundaries of the delayed region were selected for each material independently via an iterative process to maximize separation in the \TTT vs. \LO distribution of the neutron and gamma-ray events. The same boundaries were used for all solution-grown stilbene detectors, and different boundaries were used for each of the melt-grown stilbene, liquid, and plastic detectors. The light output in each detector was calibrated using a $^{22}$Na gamma-ray source. 

Neutron events were selected using a pulse shape discrimination cutoff based on the \TTT vs. \LO distribution. The light output spectrum for neutron events was produced and the following fit function was applied. This fit function approximates the detector response to a monoenergetic neutron-proton scattering interaction subject to nonlinearity in light output and detector resolution.
\begin{equation}\label{Eqn:LOFitFxn}
f(L)=\frac{mL+b}{2}\left[1-\erf\left(\frac{L-\hat{L}}{\sigma\sqrt{2}}\right)\right]-\frac{m\sigma}{\sqrt{2\pi}}e^{\frac{-(L-\hat{L})^2}{2\sigma^2}}
\end{equation}
\LOave represents the expected light output produced by a full energy proton recoil. Events with light output in the range \LOaverange were selected as full-energy proton recoils, widening the range of proton recoil directions to events within a given angle of the forward direction. For this measurement system, this angle was approximately 12$^\circ$ in the stilbene and plastic detectors and approximately 15$^\circ$ in the liquid detector. A distribution of \TTT values for the full energy events was produced, and a Gaussian fit function was applied to calculate the centroid \TTTave value which represents the expected pulse shape parameter produced by a full energy proton recoil.
 
Because the crystal axes directions are not known in several samples, the proton recoil direction is expressed in three-dimensional space with respect to an arbitrary set of axes in ($\theta$, $\phi$) spherical coordinates. $\theta$ represents the angle between the proton recoil direction and the positive $z$-axis, and $\phi$ represents the angle between the positive $x$-axis and the projection of the proton recoil direction on the $xy$-plane. Due to crystal symmetry, only one hemisphere worth of interaction directions was measured in each material. The arbitrary crystal axes are established separately for each sample, so the ($\theta$, $\phi$) coordinates do not correspond to the same crystal axes directions between measurements. The arbitrary axes were oriented so that interesting features on each distribution are away from the edges. 

\section{Anisotropy Measurements on Stilbene}\label{sec:stilbene}

\begin{figure}[!t]
	\centering
	\subfloat[Bare solution-grown cubic stilbene with crystal axes indicated.]{\includegraphics[trim={0cm 0cm 0cm 0cm},clip,width=1.25in]{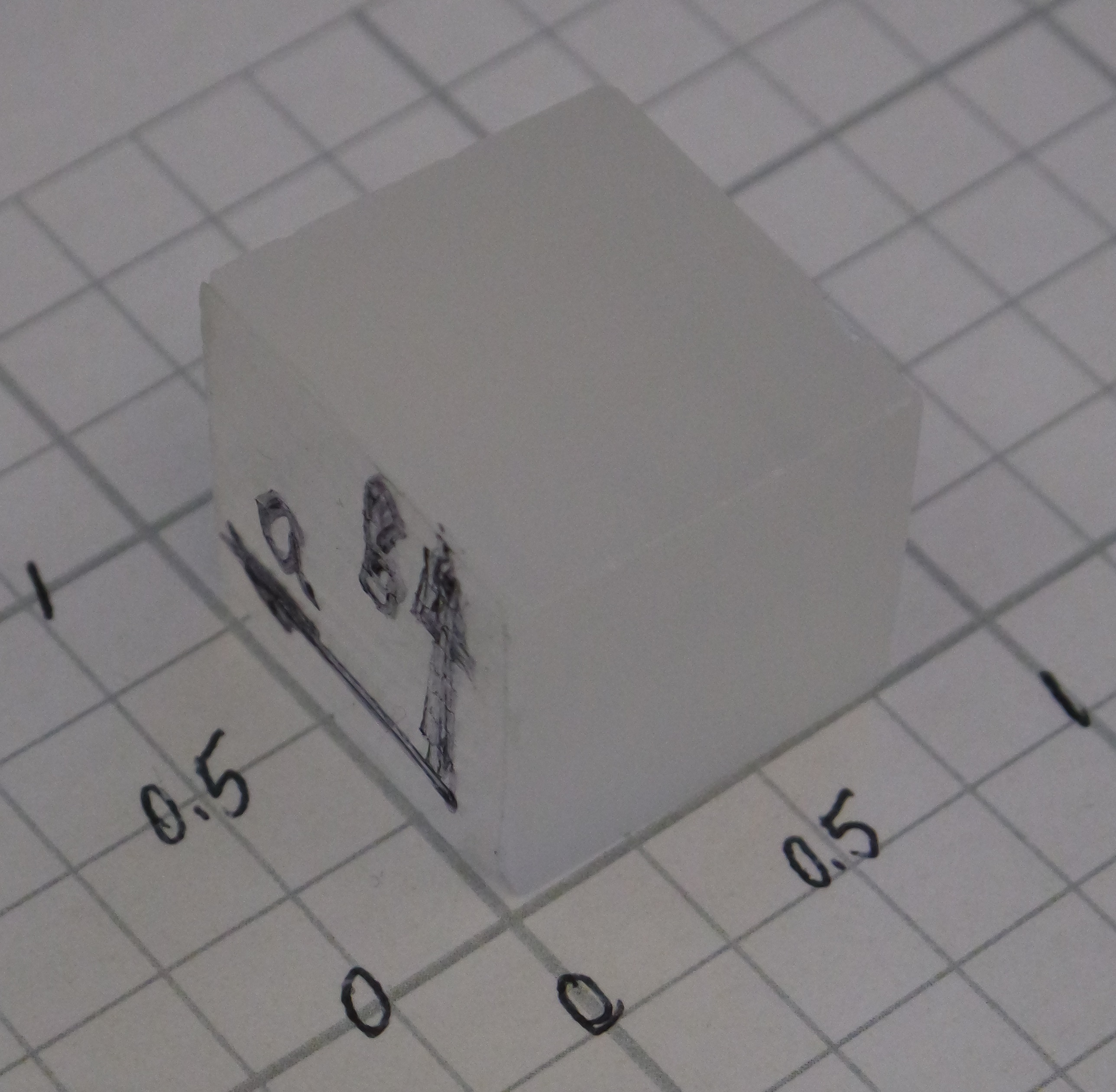}%
		\label{fig:stil_316B}}
	\hfil
	\subfloat[Melt-grown cylindrical stilbene wrapped in teflon tape.]{\includegraphics[trim={0cm 0cm 0cm 0cm},clip,width=1.25in]{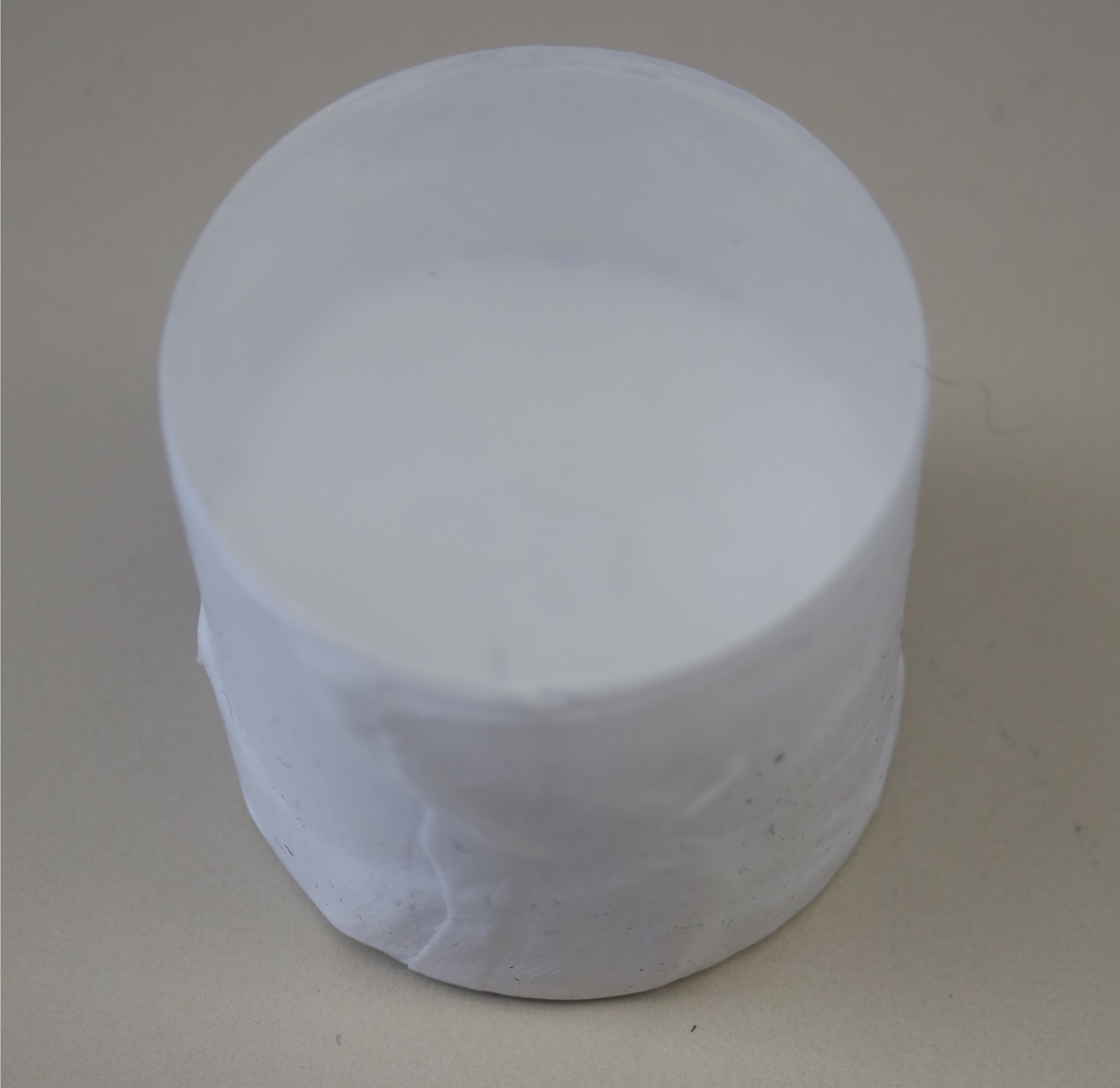}%
		\label{fig:stil_melt}}
	\caption{Photos of solution-grown and melt-grown stilbene samples.}
	\label{fig:stil}
\end{figure}

The directional response to \MeV{14.1} and \MeV{2.5} proton recoils was characterized in four stilbene samples. Three were produced using the solution-growth method. Of these three samples, two are cubic samples with \cm{1.5} edge lengths and known crystal axes directions, shown in \fig{fig:stil_316B}. The third is a rectangular prism with approximate dimensions \cm{4.3} x \cm{2.5} x \cm{0.9}. The crystal axes directions in the third sample are unknown. The fourth sample is an older stilbene crystal with unknown history and considerable wear, shown in \fig{fig:stil_melt}. It was produced using a melt growth technique, so it is unlikely to be a perfect monocrystal. The crystal is cloudy and the surface has been polished numerous times. This crystal is a cylinder of size approximately \cm{1} height and \cm{1} diameter. 

Each crystal was polished and wrapped in teflon tape before mounting to the face of a 60 mm Hamamatsu H1949-50 photomultiplier tube (PMT) assembly using V-788 optical grease. A plastic sleeve was placed over the crystal and wrapped in black tape to block out external light. The coupling of the crystal and PMT were fixed for all measurements to control the light collection efficiency.

\begin{figure}[!h]
	\centering
	\subfloat[Light output \LOave (keVee).]{
		\iftoggle{color}
		{\includegraphics[scale=1]{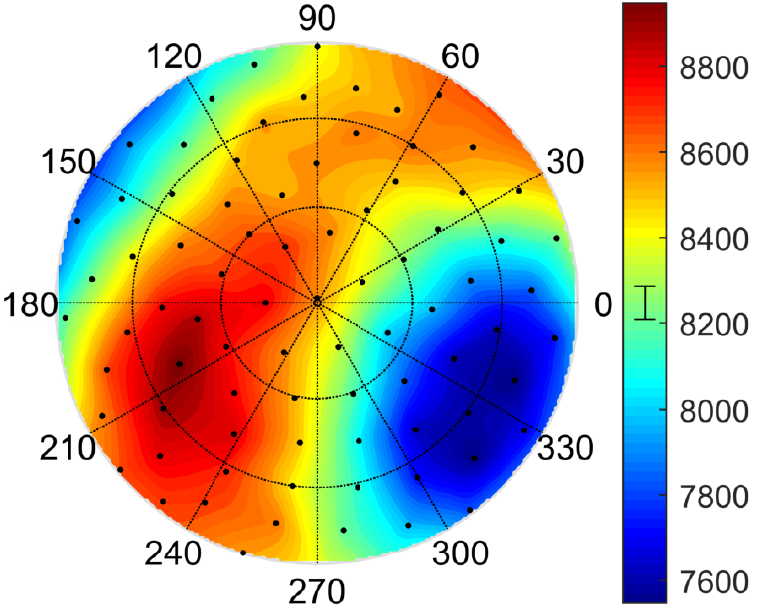}}
		{\includegraphics[scale=1]{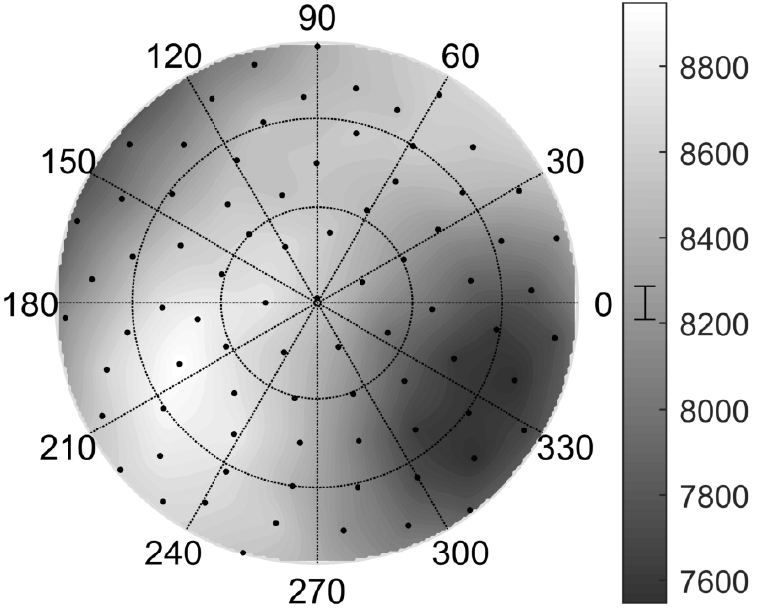}}
		\label{fig:stil_316B_DT_PP_L}}
	\hfil
	\subfloat[Pulse shape parameter \TTTave.]{
		\iftoggle{color}
		{\includegraphics[scale=1]{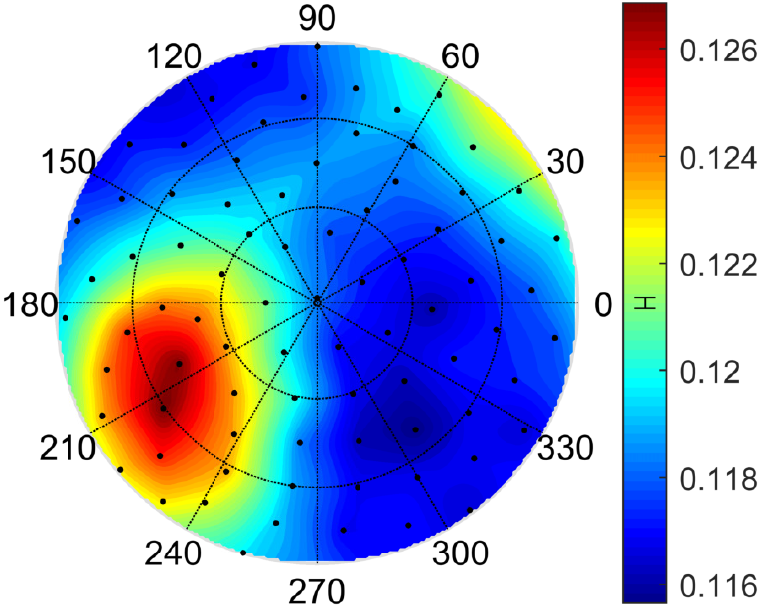}}
		{\includegraphics[scale=1]{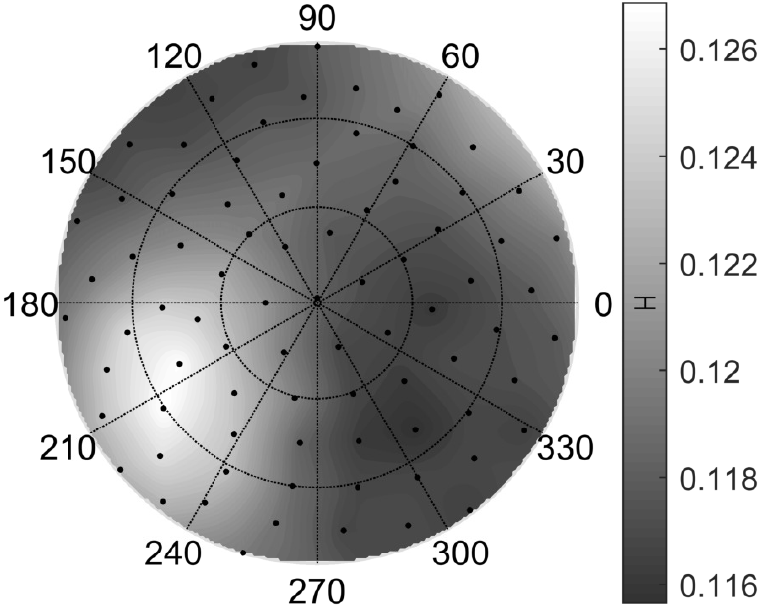}}
		\label{fig:stil_316B_DT_PP_S}}
	\caption{Response of solution-grown cubic stilbene crystal B at various recoil directions to \MeV{14.1} protons.}
	\label{fig:stil_316B_DT_PP}
\end{figure}

\fig{fig:stil_316B_DT_PP} shows the directional response of one of the cubic solution-grown stilbene samples to \MeV{14.1} proton recoil events. This data is provided in the supplementary files accompanying this paper. This visualization displays a hemisphere worth of directions essentially as if it had been flattened and viewed from above with ${r=\sqrt{1-\cos\theta}}$ increasing radially outward from 0 to 1 and $\phi$ increasing counter-clockwise from $0^\circ$ to $360^\circ$. An arbitrary set of axes has been established such that the features of interest are on the interior of the plot and not on the edges. In all such plots, the black points indicate measurements and the color scale represents a smooth interpolation between measurements. The length of the vertical bar on the colorbar indicates the average statistical error. 

The measurements on the cubic solution-grown samples with known crystal axes confirm statements by previous authors that the proton recoil direction of maximum light output is along the $b$-axis and minimum light output is along the $c'$-axis. The \LOave distribution shows a distinct maximum region at approximately $(\theta,\phi)=(50^\circ,210^\circ)$ which is confirmed to be the direction of the $b$-axis in the crystal axes. The minimum region is at approximately $(\theta,\phi)=(60^\circ,330^\circ)$ which is the $c'$-axis, and a saddle point is at approximately $(\theta,\phi)=(45^\circ,90^\circ)$ which corresponds to the $a$-axis.

The features are very similar between the \LOave and \TTTave distributions. The locations of the maximum, minimum, and saddle points are approximately the same in both distributions. This is different from anthracene, whose maximum and minimum directions are opposite in the \LOave and \TTTave distributions\cite{Schuster2016}. This is likely due to differences in molecular and crystal structure between stilbene and anthracene. One difference between the \LOave and \TTTave distributions in stilbene is that the gradient from larger to smaller values is much steeper in the \TTTave distribution, making for wider ``valleys'' in the \TTTave distribution than in the \LOave distribution. 

The qualitative features appeared the same for measurements at \MeV{2.5} and on the other two solution-grown stilbene samples at \MeV{14.1} and \MeV{2.5}. Thus, these measurements showed that the qualitative features observed in the directional distribution of the light output and pulse shape are consistent across solution-grown stilbene samples with different geometries and at different proton recoil energies. 

\begin{figure}[!h]
	\centering
	\subfloat[Light output \LOave (keVee).]{
		\iftoggle{color}
		{\includegraphics[scale=1]{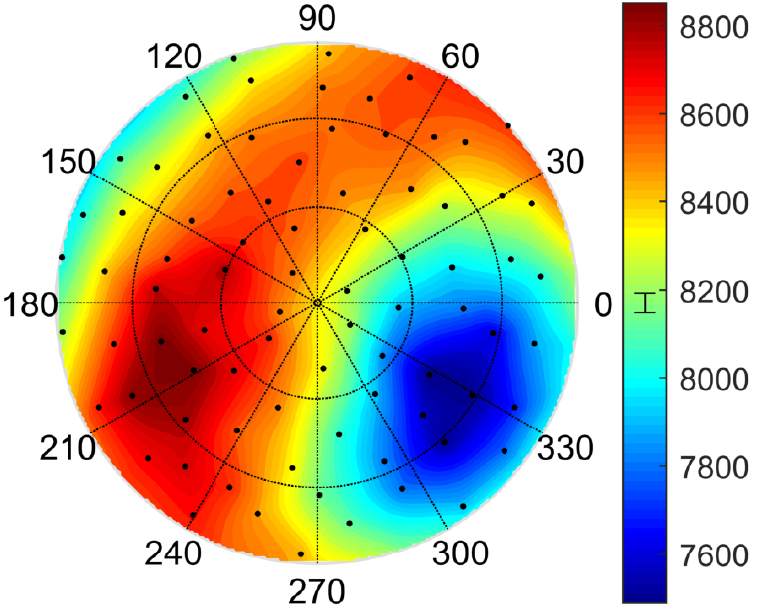}}
		{\includegraphics[scale=1]{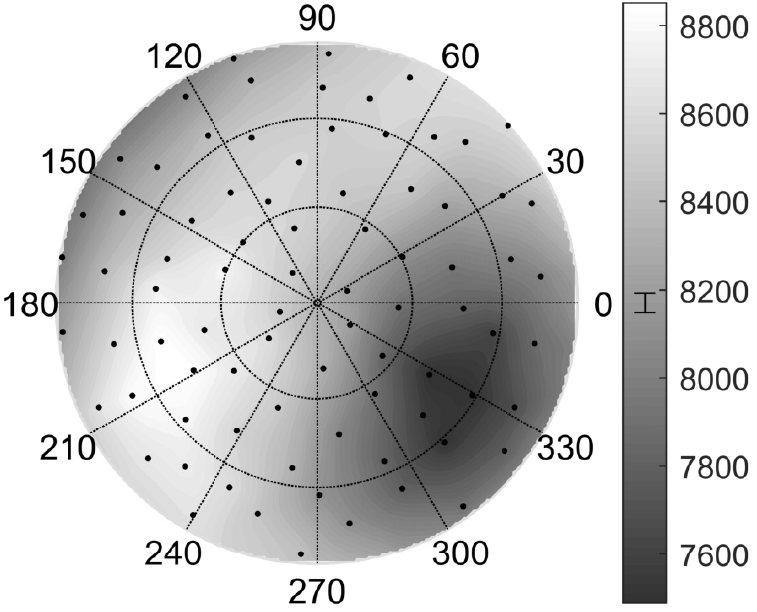}}
		\label{fig:stil_melt_DT_PP_L}}
	\hfil
	\subfloat[Pulse shape parameter \TTTave.]{
		\iftoggle{color}
		{\includegraphics[scale=1]{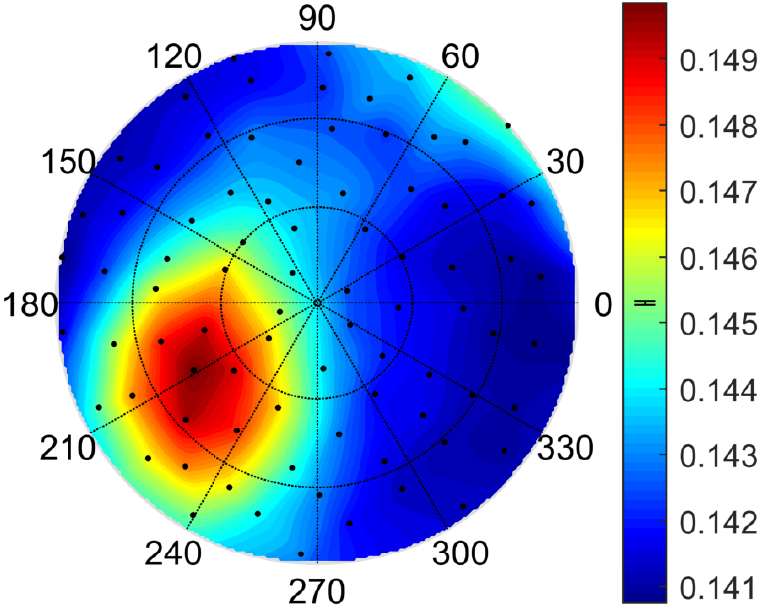}}
		{\includegraphics[scale=1]{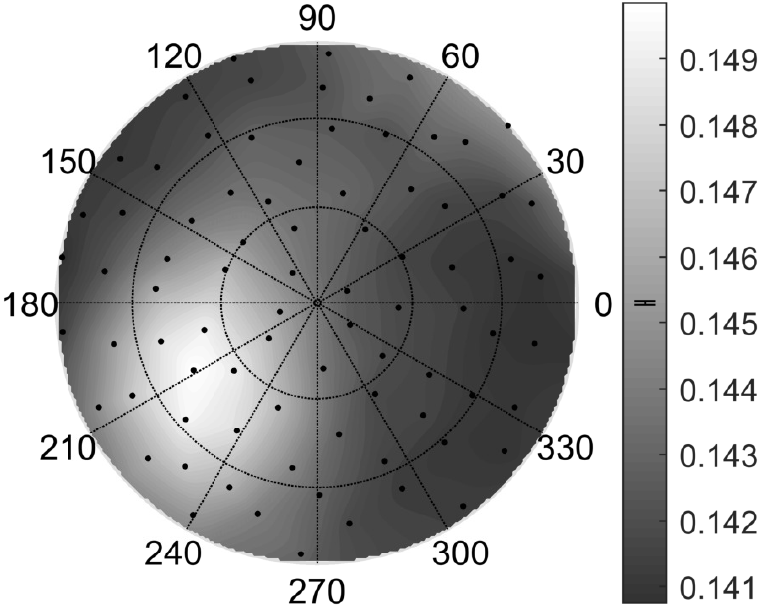}}
		\label{fig:stil_melt_DT_PP_S}}
	\caption{Response of melt-grown stilbene crystal at various recoil directions to \MeV{14.1} protons.}
	\label{fig:stil_melt_DT_PP}
\end{figure}

\fig{fig:stil_melt_DT_PP} shows the directional distributions for the melt-grown stilbene detector. The arbitrary crystal axes in the melt-grown stilbene have been oriented so that the features approximately line up with the corresponding features in the solution-grown stilbene. These distributions show that the qualitative features in the directional light output and pulse shape responses are consistent between a melt-grown and solution-grown stilbene sample.

To quantify the magnitude of the directional dependence, two metrics are calculated for each detector. The first metric is the magnitude of change across all measurements, calculated as the ratio of the maximum to minimum \LOave and \TTTave values:
\begin{align*}
\ALO&=\frac{\LOave_{\mathrm{max}}}{\LOave_{\mathrm{min}}}&
\ATTT&=\frac{\TTTave_{\mathrm{max}}}{\TTTave_{\mathrm{min}}}
\end{align*}

\tab{tab:stilbene_anis} shows the \ALO and \ATTT values and their statistical errors calculated for the four stilbene materials measured. The \ALO and \ATTT values are on the same order across all materials at \MeV{14.1}, although the \ATTT value is not directly comparable between the solution-grown materials and the melt-grown materials because the timing windows for defining the delayed region in calculating \TTT are different. The statistical error is propagated from the error produced by the fit function, and does not account for systematic errors such as fluctuations due to temperature effects or the different distribution of angles measured in each material.

\begin{table*}[htbp]
	\centering
	\caption{Magnitude of change in \LOave and \TTTave values measured for stilbene detectors. *Indicates only maximum and minimum angles from DT measurements were measured at DD energies.}
	\label{tab:stilbene_anis}
	\begin{tabular}{c|c|c|c|c|c}
		&    & \multicolumn{2}{c}{\ALO} & \multicolumn{2}{c}{\ATTT} \\
		&     & 14.1 MeV & 2.5 MeV & 14.1 MeV & 2.5 MeV \\
		\hline
		\multirow{3}[0]{*}{Solution-grown} & Cubic A & 1.200 $\pm$ 0.007 & 1.468 $\pm$ 0.050* & 1.071 $\pm$ 0.001 & 1.034 $\pm$ 0.005* \\
		& Cubic B & 1.191 $\pm$ 0.008 & 1.365 $\pm$ 0.039 & 1.100 $\pm$ 0.001 & 1.058 $\pm$ 0.007 \\
		& Rectangular & 1.191 $\pm$ 0.004 & 1.327 $\pm$ 0.026* & 1.078 $\pm$ 0.001 & 1.039 $\pm$ 0.005* \\
		\hline
		\multicolumn{2}{c}{Melt-grown}               & 1.187 $\pm$ 0.005 & 1.401 $\pm$ 0.055 & 1.066 $\pm$ 0.001 & 1.048 $\pm$ 0.003 \\
	\end{tabular}%
\end{table*}%

\begin{figure}[!t]
	\centering
	\subfloat[DT neutrons.]{\includegraphics[trim={0cm 0cm 0cm 0cm},clip,width=2.5in]{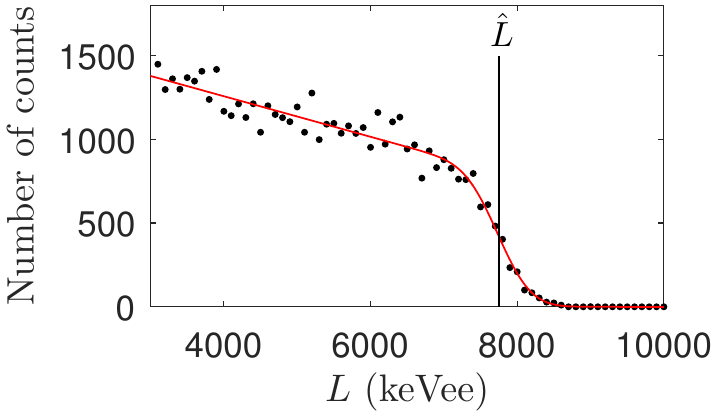}%
		\label{fig:LO_DT}}
	\hfil
	\subfloat[DD neutrons.]{\includegraphics[trim={0cm 0cm 0cm 0cm},clip,width=2.5in]{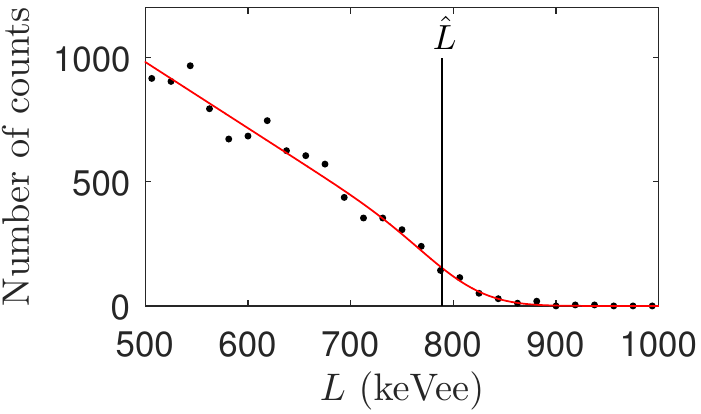}%
		\label{fig:LO_DD}}
	\caption{Light output spectrum fit as measured for DT and DD neutrons incident on the cubic solution-grown stilbene sample B.}
	\label{fig:LO}
\end{figure}

\ALO and \ATTT vary more for measurements at \MeV{2.5} than at \MeV{14.1}, likely due to the difficulty in fitting the light output spectrum at \MeV{2.5}. \fig{fig:LO} shows the neutron light output spectra for DT and DD neutrons measured at a given angle in the solution-grown cubic stilbene B sample. The edge in the light output spectrum is much clearer to the naked eye in the DT distribution than in the DD distribution for several reasons. First, the DD measurements suffer from fewer events as the cross section for the DD reaction is much lower than that for the DT reaction. Also, there are fewer optical photons produced per DD event, making for lower energy resolution due to photostatistics. The fit function is much more robust for the DT measurements, and consistently locates the same \LOave value regardless of bin size. Although the fit function locates \LOave in the DD light output spectrum, it is not a very robust process as changing the binning or light output range can change the final \LOave value significantly. In the DD measurements on the cubic B and melt-grown samples, over 70 proton recoil directions were measured, and the distribution of \LOave and \TTTave values showed features similar to the DT distributions, lending confidence that the fit function selected a consistent feature on the light output spectra. In the DD measurements on the cubic A and rectangular samples, however, measurements were only taken at the angles of maximum and minimum light output from the corresponding DT measurements so no assessment of the overall features could be made, giving less confidence that the fit function performed consistently on those measurements, so the \ALO and \ATTT values calculated from those two measurements are considered less reliable. Those values are indicated with an asterisk (*) in \tab{tab:stilbene_anis}.

\begin{figure}
	\centering
	\includegraphics[scale=.8]{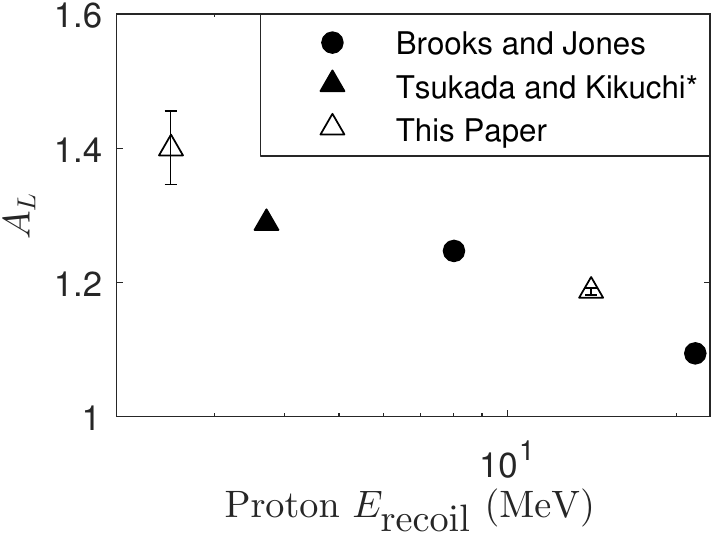}
	\caption{Magnitude of change in light output for proton recoil events at energies from 2.5-22 MeV in melt grown stilbene detectors as reported by Tsukada and Kikuchi~\cite{Tsukada1962}, Brooks and Jones~\cite{Brooks1974}, and in this work. *\ALO reported in~\cite{Tsukada1962} is for a polycrystalline sample, so it is likely lower than the true value.
		\label{fig:ALvsenergy}
	}
\end{figure}

The magnitude of change of the light output in melt-grown stilbene samples has been reported by previous authors~\cite{Tsukada1962,Brooks1974}. \fig{fig:ALvsenergy} shows the \ALO value measured by previous authors and in this paper at different proton recoil energies. All measurements are consistent with the trend that the magnitude of change in the light output decreases as the proton recoil energy increases. Comparing these results to similar measurements of anthracene made with the same detection system~\cite{Schuster2016} confirms previous authors' statements that the magnitude of the light output anisotropy in stilbene is on the same order as that in anthracene, but the pulse shape anisotropy is much greater in anthracene than in stilbene~\cite{Brooks1974}.

Another metric for quantifying the anisotropy is the variability introduced to the \LOave and \TTTave values by the effect. If the distribution of directions measured is approximated to be evenly distributed over the hemisphere, the observed standard deviation \sobs of measured \LOave and \TTTave values is related to the variability \sanis introduced by the anisotropy effect. The contribution from statistical variance is subtracted in quadrature:
\begin{equation*}
\sanis=\sqrt{\sobs^{2} - \sstat^{2}},
\end{equation*}
where $\sstat^{2}$ is the average statistical variance from the set of measurements at different recoil directions. In all cases, the correction for \sstat resulted in a \textless5\% difference between \sobs and \sanis. Values of \sanis for the solution-grown cubic B and melt-grown stilbene are given in \tab{tab:stilbene_var}, normalized to the average measured value $\mu$.

\begin{table}[htbp]
	\centering
	\caption{Normalized variability, $\sanis/\mu$, introduced by the anisotropy in directional measurements of solution-grown cubic B and melt-grown stilbene samples.}
	\begin{tabular}{c|r|r|r|r}
		                 & \multicolumn{2}{|c|}{Solution-grown} & \multicolumn{2}{c}{Melt-grown}\\
		\Erecoil (MeV)   & 14.1     & 2.5                       & 14.1     & 2.5 \\
		\hline
		\LOave           & 4.44\%   & 7.42\%                    & 4.27\%   & 7.80\% \\
		\hline
		\TTTave          & 2.26\% 	& 1.37\%                    & 1.67\%   & 0.99\% \\
	\end{tabular}%
	\label{tab:stilbene_var}%
\end{table}%

These measurements show that the variability introduced by the anisotropy at \MeV{14.1} and \MeV{2.5} is on the same order in solution-grown and melt-grown stilbene samples. Again, comparing the variability in the pulse shape is not perfectly fair as different timing windows were used for calculating \TTT in the solution-grown and melt-grown detectors. Regardless, this serves as an approximate comparison of the variability in the pulse shape as calculated for optimal neutron-gamma PSD in these materials. 


\section{Measurement of Plastic and Liquid Scintillators}
Directional measurements were also made on plastic and liquid scintillator detectors. The plastic sample was a PSD-capable plastic cylinder approximately \cm{1.25} tall and \cm{2.5} in diameter. The plastic sample was wrapped and coupled to the PMT using the same procedure as described for the stilbene detectors. The liquid scintillator was EJ-309 in a \cm{5} tall \cm{5} diameter cylinder. 

\begin{figure}[!h]
	\centering
	\subfloat[Light output \LOave (keVee).]{
		\iftoggle{color}
		{\includegraphics[scale=1]{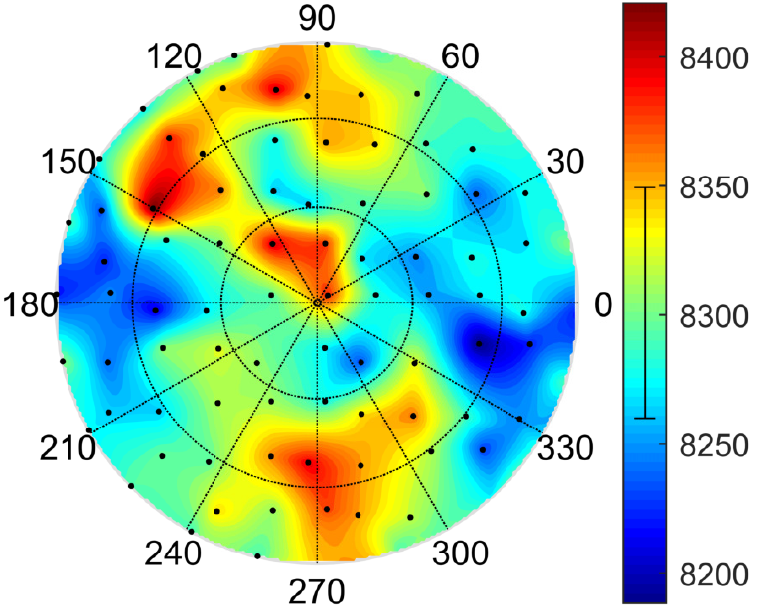}}
		{\includegraphics[scale=1]{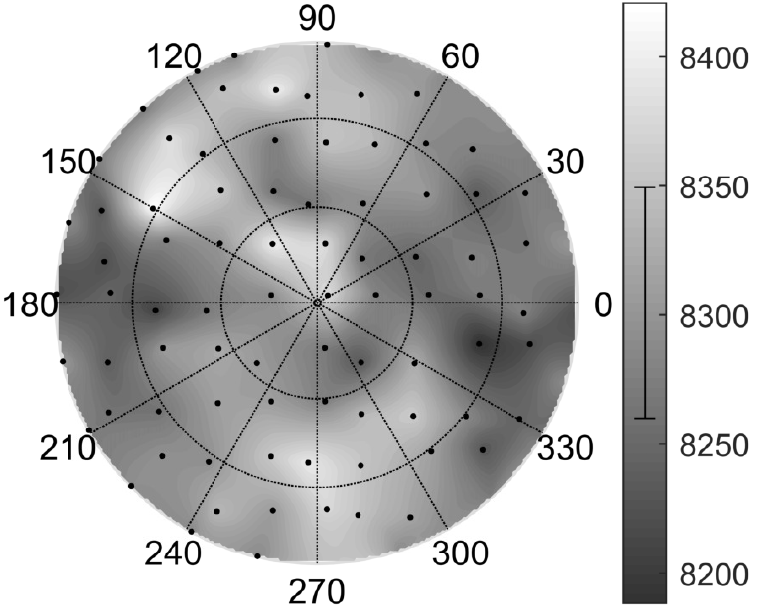}}
		\label{fig:ej309_DT_PP_L}}
	\hfil
	\subfloat[Pulse shape parameter \TTTave.]{
		\iftoggle{color}
		{\includegraphics[scale=1]{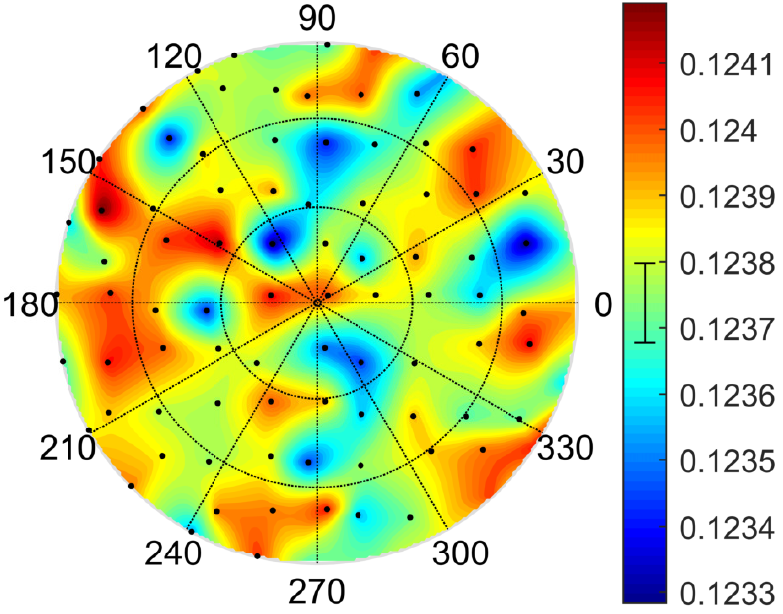}}
		{\includegraphics[scale=1]{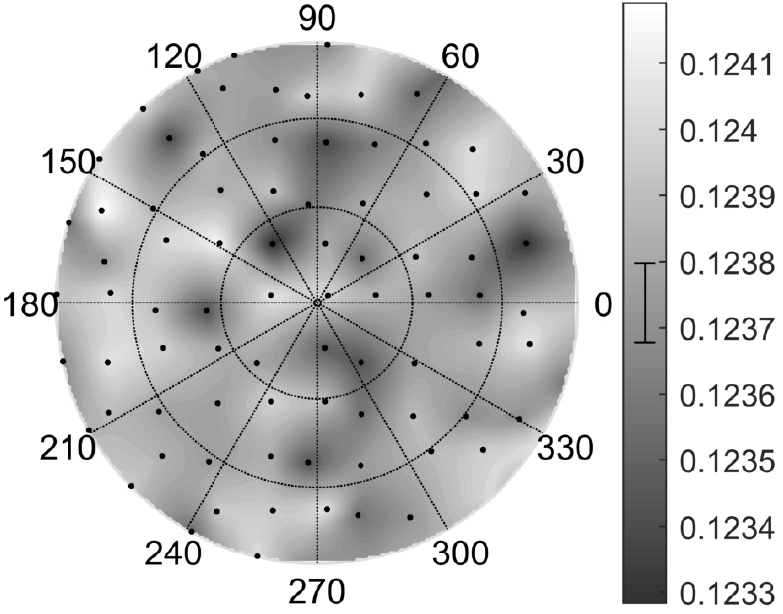}}
		\label{fig:ej309_DT_PP_S}}
	\caption{Response of EJ309 liquid at various recoil directions to \MeV{14.1} protons.}
	\label{fig:ej309_DT_PP}
\end{figure}

\begin{figure}[!h]
	\centering
	\subfloat[Light output \LOave (keVee).]{
		\iftoggle{color}
		{\includegraphics[scale=1]{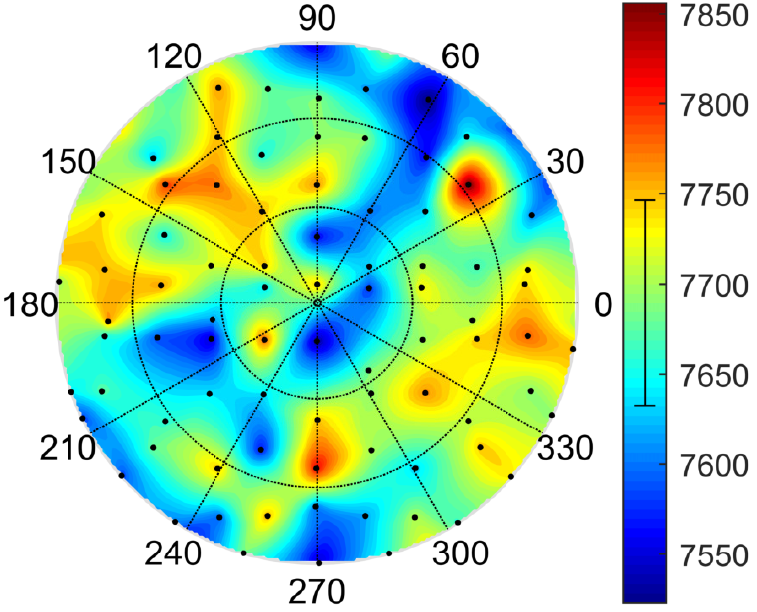}}
		{\includegraphics[scale=1]{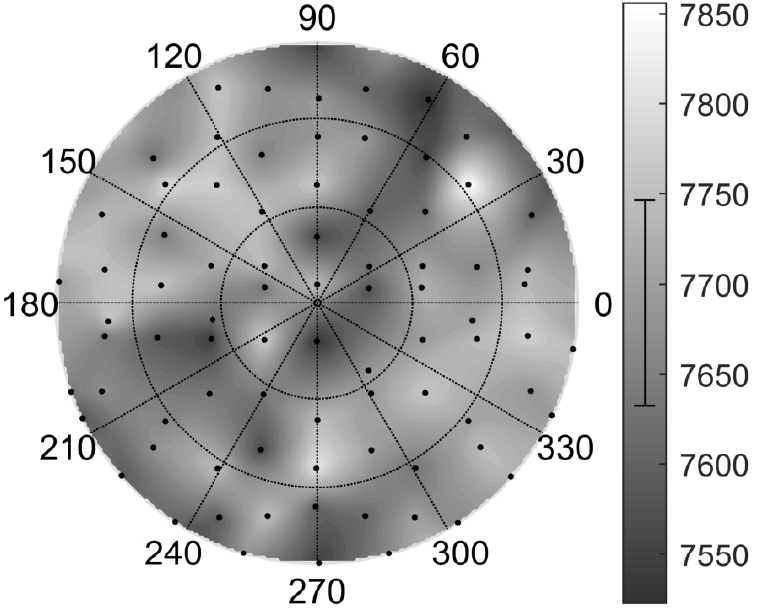}}
		\label{fig:plas_DT_PP_L}}
	\hfil
	\subfloat[Pulse shape parameter \TTTave.]{
		\iftoggle{color}
		{\includegraphics[scale=1]{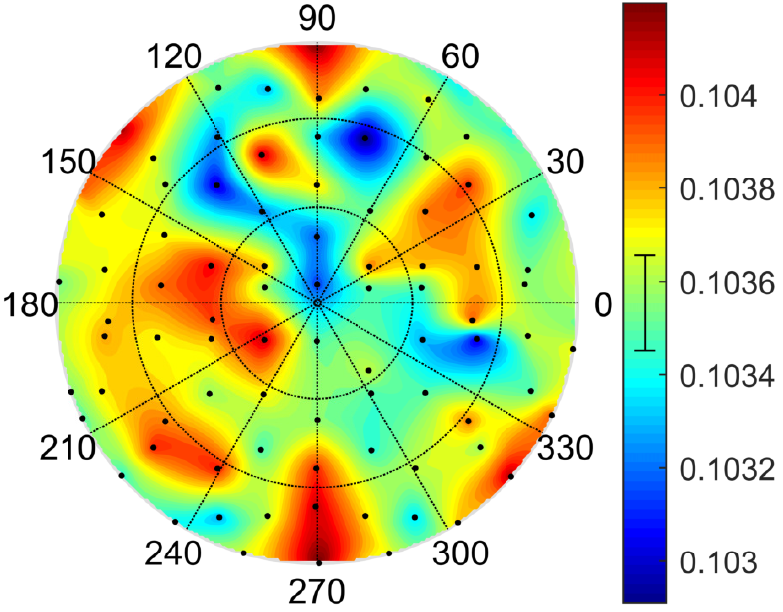}}
		{\includegraphics[scale=1]{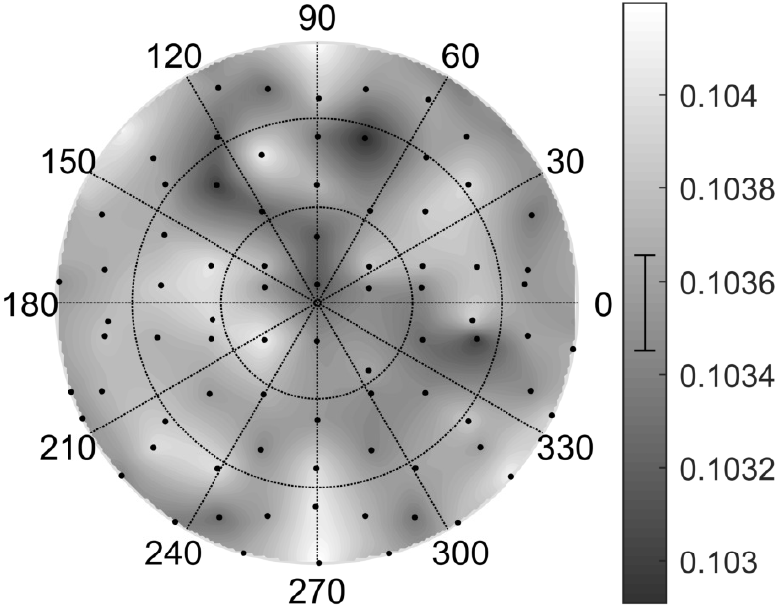}}
		\label{fig:plas_DT_PP_S}}
	\caption{Response of plastic scintillator at various recoil directions to \MeV{14.1} protons.}
	\label{fig:plas_DT_PP}
\end{figure}

The same process was used for calculating \LOave and \TTTave as was used for the stilbene measurements. \fig{fig:ej309_DT_PP} shows the directional response of the EJ309 liquid to \MeV{14.1} protons. \fig{fig:plas_DT_PP} shows the same measurements on the PSD-capable plastic scintillator. Additional measurements were taken of \MeV{2.5} proton recoils, and similar features were observed in the directional responses. 

These measurements do not show any directional dependence correlated to detector orientation. The strongest evidence supporting that conclusion is the lack of qualitative features that are observed in the stilbene detectors. These measurements do not show a maximum, minimum, and saddle point region in the hemisphere with smooth transitions in between as was observed in the stilbene measurements. Instead, the directional dependence appears to be random variability from statistical and other non-statistical effects. \tab{tab:plas_var} shows the observed variability \sobs and the average statistical variability \sstat in the liquid and plastic measurements. For all cases, \sobs is on the order of \sstat, and in the case of the light output at \MeV{2.5}, \sstat exceeds \sobs. This indicates that statistical fluctuations dominate the observed variability, though other sources may exist including fluctuations in the room temperature. The room temperature varied within a range of $1^\circ$C for the DT measurements and within $2^\circ$C for the DD measurements. If the relationship between the light output, pulse shape, and temperature were known, temperature effects could be corrected for, but no such relationship is available for this system.

Thus, both qualitative and quantitative observations confirm statements made by previous authors that no directional dependence exists in liquid and plastic organic scintillators. The lack of directional variability correlated to detector orientation in these amorphous materials indicates that the directional dependence observed in crystalline materials is in fact due to an internal effect and not produced by the measurement system or external factors. 

\begin{table}[htbp]
	\centering
	\caption{Variability in directional measurements of liquid and plastic scintillators.}
	\begin{tabular}{c|c|r|r|r|r}
		& \Erecoil              & \multicolumn{2}{|c|}{EJ309 Liquid} & \multicolumn{2}{c}{Plastic}\\
		& (MeV)                                    & 14.1   & 2.5    & 14.1   & 2.5 \\
		\hline
		\multirow{3}{*}{\LOave} & \sobs$/\mu$      & 0.60\% & 1.51\% & 0.96\% & 2.37\% \\
							    & \sstat$/\mu$     & 0.55\% & 1.87\% & 0.76\% & 3.49\% \\
		\hline
		\multirow{3}{*}{\TTTave}& \sobs$/\mu$      & 0.17\% & 0.62\% & 0.27\% & 1.10\% \\
								& \sstat$/\mu$     & 0.05\% & 0.19\% & 0.10\% & 0.39\% \\
		\hline
	\end{tabular}%
	\label{tab:plas_var}%
\end{table}%

\section{Conclusion}
The measurements reported in this paper characterize the directional dependence of the scintillation produced by \MeV{14.1} and \MeV{2.5} proton recoil events across a hemisphere worth of directions in melt-grown and, for the first time, solution-grown stilbene detectors. The light output and pulse shape produced by \MeV{14.1} and \MeV{2.5} proton recoils at a hemisphere worth of directions in a solution-grown stilbene sample is provided in the supplementary files accompanying this paper. The scintillation was confirmed to be maximal for proton recoils along the $b$ crystal axis and minimal along the $c'$ axis. The magnitude of change in the light output and pulse shape was found to be consistent across four stilbene detectors of different geometries and growth methods. The relationship between magnitude of change in light output and proton recoil energy agreed with measurements by other authors at different proton recoil energies. 

The directional response of liquid and plastic scintillator detectors was also measured and results were provided to demonstrate that no anisotropy was present. Those measurements support the leading hypothesis that the scintillation anisotropy in crystalline materials is caused by an internal effect and is not the result of an external effect on the measurement system.

\section{Acknowledgements}

The authors wish to thank John Steele for his assistance in building the motor driven rotational stage, and Natalia Zaitseva of Lawrence Livermore National Laboratory for providing the solution-grown stilbene and plastic samples.

This material is based upon work supported by the National Science Foundation Graduate Research Fellowship Program under Grant No. DGE 1106400. This material is based upon work supported by the Department of Energy National Nuclear Security Administration under Award Number DE-NA0000979 through the Nuclear Science and Security Consortium. Sandia National Laboratories is a multi-program laboratory managed and operated by Sandia Corporation, a wholly owned subsidiary of Lockheed Martin Corporation, for the U.S. Department of Energy's National Nuclear Security Administration under contract DE-AC04-94AL85000. 



\bibliographystyle{elsarticle-num} 
\bibliography{2016_05_31_references}

\end{document}